\documentclass[aps,twocolumn,showpacs,superscriptaddress]{revtex4}

\usepackage{graphicx}
\usepackage{latexsym}

\usepackage{color}

\begin{document}

\title{Dragging a polymer in a viscous fluid: steady-state and transient}

\author{Takahiro Sakaue}\thanks{Corresponding author: sakaue@phys.kyushu-u.ac.jp}
\affiliation{Department of Physics, Kyushu University 33, Fukuoka 812-8581, Japan}
\affiliation{PRESTO}

\author{Takuya Saito}
\affiliation{Department of Physics, Kyushu University 33, Fukuoka 812-8581, Japan}

\author{Hirofumi Wada}
\affiliation{Yukawa Institute for Theoretical Physics, Kyoto University, Kyoto 606-8502, Japan}

\def\Vec#1{\mbox{\boldmath $#1$}}
\def\degC{\kern-.2em\r{}\kern-.3em C}

\def\SimIneA{\hspace{0.3em}\raisebox{0.4ex}{$<$}\hspace{-0.75em}\raisebox{-.7ex}{$\sim$}\hspace{0.3em}} 

\def\SimIneB{\hspace{0.3em}\raisebox{0.4ex}{$>$}\hspace{-0.75em}\raisebox{-.7ex}{$\sim$}\hspace{0.3em}}

\date{\today}

\begin{abstract}
We study the conformation and dynamics of a single polymer chain that is pulled by a constant force applied at its one end with the other end free. 
Such a situation is relevant to the growing technology of manipulating individual macromolecules, which offers a paradigm research for probing far-from-equilibrium responses of long flexible biological polymers.
We first analyze the Rouse model for the Gaussian chains for which the exact analytical results can be obtained. 
More realistic features such as the finite extensibility, the excluded volume and the hydrodynamic interactions are taken into account with the help of the scaling argument, which leads to various nontrivial predictions such as the stretching-force-dependent friction constants. We elucidate (i) generalized dynamical equations of state describing extension/friction laws in steady-state and (ii) the tension propagation laws in the transient process. We point out that the time evolutions of the dynamic friction in the transient process crucially depend on the experimental protocol, i.e., either constant force or velocity ensemble, which might be detectable in experiments using giant DNAs and chromosomes. 
\end{abstract}


\maketitle

\section{Introduction}
\label{introduction}
A soft response to external forces is a generic property of long flexible polymers.
Take a simple example of the static stretching, where a polymer is pulled by a force $f$~\cite{deGennesBook,GrosbergBook,Macromolecules_Pincus_1976,Macromol_Marko_1995}.
Our polymer is composed of $N \gg 1$ segments with their size $a$, and its overall spatial extent is $R \simeq a N^{\nu}$ at equilibrium with the Flory exponent $\nu$.
In good solvent condition, polymers swell, i.e.,  $\nu \simeq 0.6$ due to the excluded-volume interactions, where it is well known that a linear response domain is small, and rather weak force $f \geq k_BT/R$ suffices to enter the nonlinear regime ($k_BT$ is thermal energy)~\cite{deGennesBook,GrosbergBook,Macromolecules_Pincus_1976}.
For stronger force $f \geq k_BT/a$, the inextensibility of the chain backbone introduces an additional nonlinearlity~\cite{GrosbergBook,Macromol_Marko_1995}. These aspects have been extensively studied both theoretically and experimentally.

Even more dramatic would be the dynamics, where polymers may exhibit intriguing nonequilibrium response behaviors both in a steady-state~\cite{EPL_Brochard_1993,EPL_Brochard_1995,Science_Chu_1995,PRE_Larson_Chu_1997,Macro_Netz_2008} and a transient process~\cite{EPL_Brochard_1994,Seifert_1996,PRE_Sakaue_2008,PRE_Sakaue_2010,EPL_Sebastian_2011,Vilgis_2012}.
This occurs when the rate of the external operation exceeds the inverse of the longest relaxation time in the system. As the relaxation times in the polymeric system are generally long, and their spectra are widely distributed, one expects that such a situation would be rather easily realized. The criterion for the onset of nonequilibrium response has the same scaling form as that for the significant deformation in the static response $f \geq k_BT/R$ (see Sec.~\ref{Scaling_steady}).

In the present article, we study a paradigmatic case of the dynamical response, i.e., polymers dragged by its one end.
When a constant pulling force is applied to one end of the chain at equilibrium, the chain starts to be deformed and settles in the steady-state conformation after some transient period.
We analyze this ``creep" experiment in a single chain level by two approaches.
In the first, we employ a Rouse equation of motion which provides us with exact results. The scaling approach, on the other hand, allows us to unveil nonlinear effects associated with excluded-volume and hydrodynamic interactions while leaving unknown numerical coefficients. These complementary approaches offer a comprehensive picture of the problem.

In Sec.~\ref{steady_state}, we first analyze the steady-state, and summarize the nonlinear force-extension and force-velocity relations as the generalized dynamical equation of states. In Sec.~\ref{transient}, we investigate the transient process in detail by looking at the tension propagation dynamics, where a nontrivial effect of the force magnitude depending on the dissipation mechanism is highlighted. An additional insight is provided in Sec.~\ref{Diffusion_eq}, where we formulate the transient process as the one-dimensional nonlinear diffusion process of local segment density towards the stretching direction. Finally, we summarize in Sec.~\ref{summary} with some remarks and perspectives.

\section{Steady-state properties}
\label{steady_state}
We first analyze the steady-state conformation of a flexible polymer chain that is forced to move by a constant force applied at its one end with the other end free. 
\subsection{Rouse model}
\label{Rouse_steady}
Parameters characterizing our polymer chain are the total number $N$ of bonds with equilibrium distance $a$ between neighboring beads (bond length or persistence length). We consider the Rouse dynamics in which the surrounding fluid provides a local isotropic drag proportional to the local velocity, ${\bf f}_{drag}=\gamma \dot{\bf r}$, where  the friction coefficient $\gamma$ is proportional to $\eta a$ ($\eta$ is the viscosity of solvent) and the dot represents time derivative throughout this paper.
Since the entropic elastic forces exactly balance the drag forces in the Rouse dynamics, the equation of motion is given by~\cite{DoiEdwards}
\begin{eqnarray}
\gamma \dot{\bf r}(n,t) &=& -k \frac{\partial^2{\bf r} (n,t)}{\partial n^2} + \mbox{\boldmath$\xi$}(n,t)+ {\bf f}(n).  
\label{eq:1}
\end{eqnarray}
where the spring constant is given by $k=3k_{B}T/a^2$, and the random force $\mbox{\boldmath $\xi$}$ (with its mean zero) describes the coupling to the thermal bath and satisfies
the usual fluctuation-dissipation relationship $\langle\mbox{\boldmath$\xi$}_n(t) \mbox{\boldmath$\xi$}_m(t')  \rangle = 2 \gamma k_{ B}T \delta_{n,m}\delta(t-t')$. 
The constant external force applied at its one end $s=0$ (and along $x$ direction) is explicitly described in Eq.~(\ref{eq:1}) as  
\begin{eqnarray}
\mbox{\boldmath$f$}_n = 2f \delta(n) \mbox{\boldmath$e$}_x , 
\label{f_Rouse_steady}
\end{eqnarray}
rather than appearing in the boundary conditions for ${\bf r}(n,t)$ given below. 
Note that the factor $2$ in Eq.~(\ref{f_Rouse_steady}) assures that $\int_0^{N} dn \,\mbox{\boldmath$f$} = f \mbox{\boldmath$e$}_x$.
The boundary conditions for ${\bf r}$ are thus those for the force-free at both ends,
$\partial {\bf r}(n)/\partial n|_{n=0}=\partial {\bf r}(n)/\partial n|_{n=N}=0$.

Using the standard procedure based on the normal mode analysis, various quantities can be computed analytically~\cite{DoiEdwards}.  
The center of mass position ${\bf r}_G(t)=N^{-1} \int_0^N {\bf r}(n,t)dn$ immediately gives the force-velocity ($f$-$V$) relation with the friction coefficient $\Gamma_{\rm Rouse} = N \gamma$;
\begin{eqnarray}
\Vec{V} = \langle \dot{\bf r}_G \rangle &=& \frac{f \mbox{\boldmath$e$}_x}{N \gamma}.
\label{Rouse_f-V}
\end{eqnarray}
The polymer moves at a speed $V=f/(N\gamma)$ as a whole at steady state.
The corresponding chain deformation is manifested in the higher modes, which  leads to
\begin{eqnarray}
\langle [{\bf r}(n)-{\bf r}_G] \rangle &=& \frac{Na^2 f }{k_BT}\left[ \frac{1}{9}-\frac{1}{3}\left( \frac{n}{N}\right)+\frac{1}{6}\left( \frac{n}{N}\right)^2\right]\mbox{\boldmath$e$}_x
\label{Rn}
\end{eqnarray}
From this, one can deduce the force-extension ($f$-$L$) relation
\begin{eqnarray}
L = |\langle {\bf r}(0)-{\bf r}(N)\rangle| = \frac{Na^2}{6k_{B}T}f
\label{Rouse_f-L}
\end{eqnarray}
By comparing eq.~(\ref{Rouse_f-L}) with the equilibrium coil size $R = a N^{1/2}$, one indeed find the threshold force $f = 6k_{B}T/R$ above which the deformation becomes apparent.
One can also obtain the velocity-extension ($V$-$L$) and $f$-$V$-$L$ relations by eliminating $f$ or $N$ from eq.~(\ref{Rouse_f-V}),~(\ref{Rouse_f-L}) as
\begin{equation}
L = \frac{\gamma a^2 N^2}{6k_{B}T} V = \frac{a^2}{6 k_{B}T \gamma}\frac{f^2}{V}
\label{Rouse_f-V-L}
\end{equation}
The dragging force builds up along the chain from the downstream end;
the internal tension at length $n$ in the polymer chain is along the $x$ direction, ${\bf T}(n)=f(n){\bf e}_x$, where
$f(n)=k\langle\partial {\bf r}/\partial n \rangle = f(1-n/N)$.

\begin{figure}[t]
\begin{center}
\includegraphics[width=7cm]{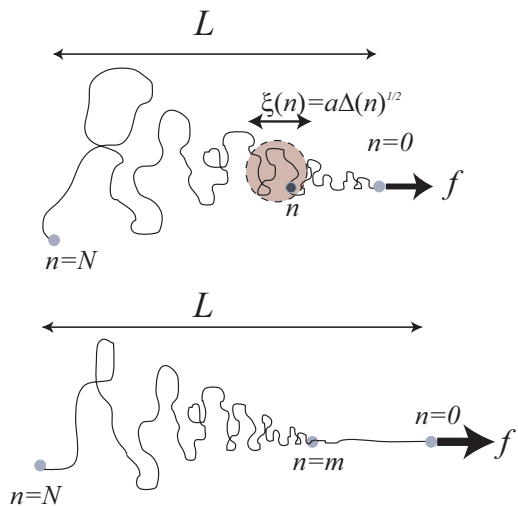}
\caption{Schematic representations of a dragged Rouse chain by a constant force $f$. The bottom picture corresponds to the case with high force $f>f_C$ where there is a stem of length $ma$ in the front region.}
\label{Rouse_schematics}
\end{center}
\end{figure}

{\it Second moment}:
Second moment quantities allow us to gain more information on how the stretching alters entropic properties in our Gaussian random coil conformations. 
A straightforward calculation yields 
\begin{eqnarray}
& & \langle |{\bf r}(n)-{\bf r}(m)|^2\rangle = a (n-m) \nonumber \\
& & +\frac{N^2a^2}{9}\left(\frac{a f}{k_{B}T}\right)^2\left[ \left(1-\frac{n+m}{2N}\right) \left(\frac{n-m}{N} \right)\right]^2.
\label{Rnm}
\end{eqnarray}
The first diffusive term describes entropic randomization effects while the second term arises from the stretching of the chain.
Comparing these two contributions leads to the characteristic size $\Delta=n-m$ (as function of $n$) separating entropy-dominated and stretching-dominated regimes.
Assuming a sufficiently long chain $N \gg 1$, we obtain, up to the first order of $\Delta /N$, $(a f(n)/k_BT)^2 \Delta \simeq 1$, where $f(n)=f(1-n/N)$ is the tension at $n$ given above.
Introducing a $n$-{\it dependent blob size} as $\xi(n)\simeq a \Delta^{1/2}$, we find the relation  
\begin{eqnarray}
\frac{f(n)\xi(n)}{3k_BT} &\simeq& 1.
\label{eq:f_c} 
\end{eqnarray}
This has a clear and important physical meaning (see Fig.~\ref{Rouse_schematics} (top)); Sitting on the position of the bead $n$, and look at the chain conformation around it. The tensile effect will be a weak perturbation and the chain conformation is essentially the same as the trajectory of a random walk at the scale smaller than $\xi(n) \simeq 3k_BT/f(n)$. At larger length scales, however, the chain is highly deformed to the stretched conformations. An asymmetric way of the driving, i.e., pulling one end, leads to the nonuniform tension profile and the size-dependent blob size $\xi(n)$.

{\it Effect of the finite chain extensibility}:
As we have seen, the tension is highest at the pulling site. From eq.~(\ref{Rn}), the distance between the dragged bead ($n=0$) and the next bead ($n=1$) is calculated as
\begin{eqnarray}
\langle |{\Vec r}_1 - {\Vec r}_0| \rangle = \frac{Na^2 f}{3k_{B}T}\left( \frac{1}{N} - \frac{1}{2N^2}\right) \simeq \frac{a^2 f}{3k_{B}T}
\end{eqnarray} 
From this, we notice that when the force reaches the threshold 
\begin{eqnarray}
f_C = \frac{3k_{B}T}{a}.
\label{f*}
\end{eqnarray}
the bond reaches its maximum extension $\langle |{\Vec r}_1 - {\Vec r}_0 |\rangle = a$. For stronger forces, the Rouse model which maintains the bond by a harmonic potential is no longer applicable. Below, we will discuss this effect of the finite chain extensibility in a simplified manner.

Consider that the chain is completely extended up to the $m$-th bead counting from the dragging site, forming a stem of the length $ma$ (see Fig.~\ref{Rouse_schematics} (bottom)). This means that the tension at the $m$-th bead is equal to $f_C$. For the remaining rear part with $N-m$ segments, the results for the Rouse model obtained above may be applied, so that using Eqs.~(\ref{Rouse_f-L}) and~(\ref{f*}), we obtain its extension $\simeq a^2(N-m)f_C/6k_BT=(N-m)a/2$. 
Since the extension of the stem is $ma$, the total end-to-end distance of this stem-flower shape is $L=(N+m)a/2$.
The number $m$ can be determined from the total force balance 
$f=N\gamma V =  f_C +m \gamma V$, leading to $m=(f-f_C)/(\gamma V) = N(1-3 k_{B}T/fa)$.
The force-extension relation in this strong force regime is 
\begin{eqnarray}
L &=& Na \left[ 1- \frac{3}{2} \left( \frac{k_{B}T}{fa}\right)\right]
\label{f-L_stem_flower}
\end{eqnarray}

\begin{figure}[h]
\begin{center}
\includegraphics[width=8cm]{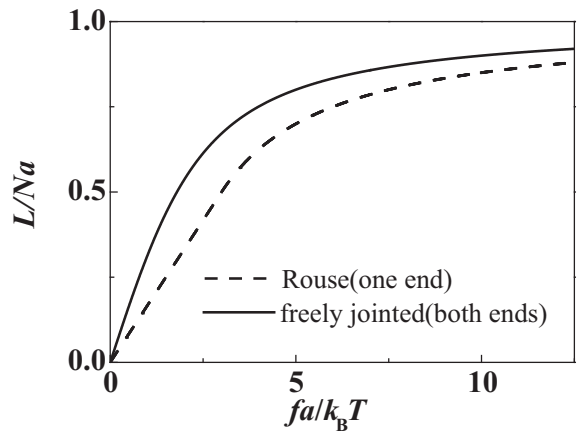}
\caption{Force-extension relations of a dragged Rouse chain pulled by its one end and a free-jointed chain pulled by its two ends. }
\label{f-L_Rouse_FJC}
\end{center}
\end{figure}
It is instructive to compare these results eq.~(\ref{Rouse_f-L}) and~(\ref{f-L_stem_flower}) with the force-extension relation for the free-jointed chain pulled by its both ends, i.e., a static stretching (Fig.~\ref{f-L_Rouse_FJC}). In this case, the tension thus the conformation is uniform along the chain, and the exact force-extension in the full force range is available~\cite{GrosbergBook}:
\begin{eqnarray}
L
&=&Na \left[ \coth{\left( \frac{fa}{k_BT}\right)} - \frac{k_BT}{fa}\right]
\nonumber \\
&=&\left\{
           \begin{array}{ll}
              (a^2/3k_BT) Nf &   \qquad( f  \ll k_BT/a) \\
              Na\left[1-k_BT/(fa) \right] &  \qquad (f \gg k_BT/a) 
           \end{array}
        \right.
\end{eqnarray}
Recall that a freely-jointed chain and a Rouse chain are similar in the sense both have no excluded volume effect, but the former is now in the static situation pulled by both ends while the latter is in the dynamical steady state with strong inhomogeneity along the chain. Nonetheless, we see that there is no distinction between these two in the scaling level. Difference, however, clearly exists in the exact prefactors, i.e., the one-site dragging results in less extension in the same force as is expected. 

We have adopted the simplest prescription to analyze the finite chain extensibility effect at high force $f>f_C$. In general, however, more sophisticated approaches require the specification of the flexibility mechanism in the segment scale. This can be, for instance, a freely-jointed model or a work-like-chain model which is suitable for the chain with a uniform bending elasticity.
The present simple prescription, however, suffices for the investigation of scaling properties. 
In the following scaling approach, we shall see that both cases ($f< f_C$ and $f > f_C$) can be treated in the same formalism, but with different characteristic scaling exponents $\alpha$ and $\beta$ defined below (eqs.~(\ref{f-L-V_2})-(\ref{f-L-N})).

\subsection{Scaling approach}
\label{Scaling_steady}
The Rouse model is applicable to polymers in melts~\cite{DoiEdwards}. But in many of other situations, the excluded-volume interactions and/or the solvent mediated hydrodynamic interactions should be taken into account. The former gives rise to long range correlation along the chain and the latter makes the friction coefficient conformation dependent.
The resultant nonlinear effects can be treated within the framework of the scaling theory.

{\it Onset of the strong deformation}:
Assume that a pulling force is weak enough so that the dragged chain takes an equilibrium conformation $R \simeq a N^{\nu}$ with the Flory exponent $\nu$.
We denote the friction coefficient of the chain as $\Gamma = \gamma (R/a)^{z-2}$ with the so-called dynamic exponent $z$~\cite{deGennesBook}.
A longest relaxation time of the polymer coil is then $\tau_{\rm eq} \simeq \Gamma R^2/k_BT \simeq \tau_0 (R/a)^z$, where $\tau_0 = \gamma a^2/k_BT \simeq \eta a^3/k_BT$ is a segment scale microscopic time.
Comparing this with the typical velocity gradient ${\dot \gamma} \simeq V/R \simeq (R/a)^{1-z}  f/(a \gamma)$, one finds a characteristic force $f \simeq k_BT/R$ above which the chain exhibits substantial deformations.
Note that this coincides with the static criterion given by comparing sizes of the Pincus blob $k_B T/f$ and the equilibrium coil $R$~\cite{deGennesBook,GrosbergBook,Macromolecules_Pincus_1976}.
For weaker force, the chain is assumed to be in near equilibrium where ordinary linear response applies.

{\it Steady-state conformation}:
The conformation of a tethered chain submitted to a solvent flow of velocity $V$ was studied by Brochard-Wyart using scaling approach~\cite{EPL_Brochard_1993,EPL_Brochard_1995}.
Here we present its generalized form using two critical exponents $\nu$ and $z$.
Let us assume that the force is moderately strong, i.e., $k_BT/R < f < k_BT/a \simeq f_C$.
As we have already seen in the analysis of Rouse model, for a chain dragged in a viscous fluid, there is a blob size $\xi(x)$ below which the effect of the pulling is insignificant, and an equilibrium formula for the Flory relation $\xi(x) \simeq a g(x)^{\nu}$ can be applied, where $g$ represents the number of segments in the blob.
Note that the one-dimensional coordinate $x$ is taken along the long axis of a stretched polymer chain (whose shape is assumed to have an axial symmetry), and {\it not} the arc-length parameter as in the previous section.  We assume that the same scaling behavior holds for the local friction coefficient
\begin{eqnarray}
\Gamma (\xi) &=& \gamma \left(\frac{\xi}{a}\right)^{z-2}
\label{eq:Gamma}
\end{eqnarray}
Hydrodynamic interaction among beads within a blob can be taken into account by setting $z=3$ (non-draining case), leading to the Stokes formula $\Gamma \simeq \eta \xi$. On the other hand, setting $z=(1+2\nu)/\nu$ amounts to assume that the solvent is just immobile (free-draining). We then recover the Rouse-type friction law $\Gamma(\xi) \simeq \eta a g$.
The overall deformed conformation can be pictured as a sequence of blobs of size $\xi$ which are hydrodynamically decoupled.
Hence, the hydrodynamic friction force acting at the position $x$ can be evaluated as an integral of the contribution which builds up from the downstream free end;
\begin{eqnarray}
f_{{\rm drag}}(x) \simeq V \int_0^{x} dx \ \frac{\Gamma}{\xi}
\end{eqnarray}
Applying a local force balance equation $\xi \simeq k_BT/f_{{\rm drag}}$~\cite{Macromolecules_Pincus_1976}, we obtain the conformation profile as 
\begin{eqnarray}
\xi(x) \simeq a \left( \frac{\tau_0 V x}{a^2}\right)^{1/(2-z)}
\label{steady_profile}
\end{eqnarray}
By equating the friction force at the pulled end to the applied force (total force balance), we find a steady-state $f$-$V$-$L$ relation, i.e., a generalized version of eq.~(\ref{Rouse_f-V-L}) in the form
\begin{eqnarray}
f \simeq \frac{k_BT}{a}\left( \frac{\tau_0 V L}{a^2}\right)^{1/(z-2)}
\label{f-V-L}
\end{eqnarray}
To obtain a generalized $V$-$L$ relation, we integrate the number of segments from the downstream free end to the pulled site, and equate it with the total segment number, i.e., mass conservation law:
$N = \int_0^{L} (g/\xi) dx$, which leads to 
\begin{eqnarray}
 L &\simeq& a N^{(2-z)\nu/(1+\nu-z\nu)} \left( \frac{\tau_0 V}{a}\right)^{(\nu-1)/(1+\nu-z\nu)}
\label{V-L-N}
\end{eqnarray}
It is convenient to summarize eq.~(\ref{f-V-L}) and eq.~(\ref{V-L-N}) as the following relations which we call the dynamical equation of state;
\begin{eqnarray}
LV \simeq \frac{a^2}{\tau_0} \left( \frac{f a}{k_{\rm B}T}\right)^{\alpha}
\label{f-L-V_2}
\end{eqnarray}
\begin{eqnarray}
N V \simeq \frac{a}{\tau_0}  \left(\frac{fa}{k_{\rm B}T} \right)^{\alpha - \beta}
\label{f-V-N}
\end{eqnarray}
\begin{eqnarray}
L \simeq Na \left( \frac{f a}{k_{\rm B}T}\right)^{\beta}
\label{f-L-N}
\end{eqnarray}
where defining two exponents
\begin{eqnarray}
 \alpha = z-2, \ \beta &=& (1-\nu)/\nu.
 \label{eq:beta}
\end{eqnarray}
turns out to be useful, but it should be kept in mind that these are merely transcriptions of the original basic dynamic and static exponents $z$ and $\nu$.
Note that all of these relations reduce to those of Rouse model for an ideal statistics $\nu=1/2$ in the free-draining limit $z=(1+2\nu)/\nu=4$.
However, the excluded-volume and hydrodynamic interactions alter the exponents.
In particular, one can see the nonlinear friction law of the hydrodynamic origin from eq.~(\ref{f-V-N}) and the nonlinear extension law of the excluded-volume origin from eq.~(\ref{f-L-N}), i.e., the linear $f$-$V$ or $f$-$L$ relation holds only for the free-draining condition or for the chain with $\nu=1/2$, respectively.
Note also that the scaling $f$-$L$ relation is the same as that for a well-known situation of the static chain stretching, i.e., pulling force applied to both its ends~\cite{deGennesBook,GrosbergBook,Macromolecules_Pincus_1976}. Therefore, the difference in the force-extension relation between dynamic and static stretching is manifested only in the numerical coefficient (compare eq.~(\ref{Rouse_f-L}) and~(\ref{f-L-N}) with $\nu=1/2$).
At the free end of the chain, the largest blob of size $\xi_{{\rm free}}$ there experiences the hydrodynamic dragging force acting on itself only. From the Pincus relation $k_BT/\xi_{{\rm free}} \simeq \gamma V (\xi_{{\rm free}}/a)^{z-2}$, we find
\begin{eqnarray}
\xi_{{\rm free}} \simeq a \left( \frac{\tau_0 V}{a}\right)^{1/(1-z)}
\label{xi_free}
\end{eqnarray}

{\it Effect of the finite chain extensibility}:
So far we have dealt with the ``trumpet" regime $f < f_C$. For stronger force, we have to take account of the finite chain extensibility, as we have seen in the analysis of Rouse model (Sec. \ref{Rouse_steady}), where the dragged chain takes a so-called ``stem-flower" conformation~\cite{EPL_Brochard_1995}. Our formalism is applicable even to such a situation, i.e., a set of basic equations eqs.~(\ref{f-L-V_2})-(\ref{f-L-N}) remains intact with the assignment of appropriate exponents $\alpha = 1$ and $\beta=0$. Indeed, $\beta=0$ in eq.~(\ref{f-L-N})  and $\alpha=1$ in eq.~(\ref{f-V-N}) indicate $L \sim N$ and $f \sim N V$, respectively, i.e., full stretching on the scaling level~\cite{EPL_Brochard_1995} (see eq.~(\ref{f-L_stem_flower}) for Rouse model).
The size of the largest blob $\xi_{{\rm free}}$ is still given by eq.~(\ref{xi_free}) until the point $\xi_{{\rm free}} > a \Leftrightarrow f < N k_BT/a$, where use has been made of eq.~(\ref{f-V-N}) with $\alpha=1$ and $\beta=0$. For stronger force $f > N k_BT/a$, an entropic coiling at the rear end is negligible so that the dragged chain takes almost fully stretched conformation.

\section{Transient dynamics}
\label{transient}
Now we turn our attention to the transient dynamics after the sudden action of a pulling force at one end.
Here we will observe that there are a variety of tension-propagation dynamics depending on the types of hydrodynamics and/or excluded volume interactions.

\subsection{Rouse model}
\label{transient_Rouse}
To set the stage, we again start with the analysis of the Rouse model dynamics.
The basic equations are already given in Sec.~\ref{Rouse_steady}. 
We only need to modify eq.~(\ref{f_Rouse_steady}). The dragging force is switched on at $t=0$, then,
\begin{eqnarray}
\mbox{\boldmath$f$}_n = 2 f u (t) \delta (n) \mbox{\boldmath$e$}_x,
\label{f_Rouse_transient}
\end{eqnarray}
where $u(t)=1$ for $t \geq 0$ and $u(t)=0$ for $t < 0$.
The resulting linear equations can be solved exactly again using the normal mode analysis.
We are particularly interested in the apparent pulling velocity of the polymer chain (i.e., the velocity of the pulling end),   
 ${\Vec V}(t) = \langle \dot{\bf r}_0\rangle$, which is found to be
\begin{eqnarray}
{\Vec V}(t) &=&  \frac{f }{N\gamma} \left[ 1 +  G(t) \right] {\Vec e}_x ,
\end{eqnarray}
where the propagator $G(t)$ is
\begin{eqnarray}
G(t)
&=&  2 \sum_{p=1}^\infty \exp\left( -\frac{\pi^2 k}{N^2 \gamma}p^2 t \right) \approx \left( \frac{t}{\tau_{\rm R}} \right)^{-1/2}.
\end{eqnarray}
In the last near-equality, we have replaced the summation with the integral, and the Rouse time $\tau_{\rm R} = \gamma a^2 N^2/(3 \pi k_BT)$ has been introduced~\cite{DoiEdwards}.
The final result
\begin{eqnarray}
{\Vec V}(t) \simeq \frac{f }{N \gamma} \left[ 1 +   \left( \frac{t}{\tau_{\rm R}} \right)^{-1/2} \right] \mbox{\boldmath$e$}_x
\end{eqnarray}
indicates that in early stage $t \ll \tau_{\rm R}$, the velocity decreases as ${\Vec V}(t) \simeq f/(N \gamma) (t/\tau_{\rm R})^{-1/2}\mbox{\boldmath$e$}_x$, and eventually reaches the steady-state with the velocity ${\Vec V}(t)  =f/(N \gamma) \mbox{\boldmath$e$}_x$ after the characteristic time $\tau_{\rm R}$.
Its physical interpretation is as the following. At $t < \tau_{\rm R}$, only a part of segments $M(t) < N$ close to the dragged site can respond to the pulling force, while other $N-M(t)$ segments in the rear are essentially unaffected yet, thus taking an unperturbed conformation at rest. Then, we have the force balance equation, 
$\gamma M(t) V(t) \simeq f$, from which we obtain
\begin{eqnarray}
 M(t) &\simeq& \left( \frac{t}{\tau_R} \right)^{1/2}
 \label{Rouse_t_f_M_V}
\end{eqnarray}
The tension applied at the dragged site diffusively propagates along the chain backbone.

\begin{figure}[t]
\begin{center}
\includegraphics[width=7cm]{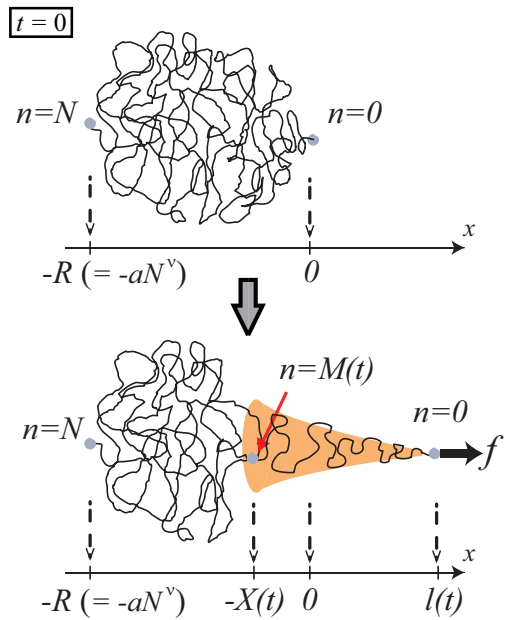}
\caption{Schematic representations of the transient dynamics in stretching process.}
\label{transient_image}
\end{center}
\end{figure}

\subsection{Scaling approach}
\label{transient_Scaling}
Imagine that one end segment located at the origin is started to be pulled to the $x$ direction at $t=0$. Before the pulling force $f$ is applied ($t<0$), the chain at rest takes the equilibrium conformation.
As already discussed in Sec.~\ref{Scaling_steady}, the whole chain would follow such a force with a velocity $V = f/\Gamma$ by retaining equilibrium conformations, if the force is below the threshold $f \simeq k_B T/R$.
However, for stronger forces, only the fraction of the chain (with the number of segment $g_0 \simeq (\xi_0/a)^{1/\nu} \simeq (fa/k_BT)^{-1/\nu}$) close to the driving site can immediately respond and set in motion, while keeping the remaining rear part at rest yet.
Such a transient response can be described by what we call the two-phase formalism originally introduced in the problem of driven translocation across a nanopore~\cite{PRE_Sakaue_2008,PRE_Sakaue_2010}, in which one analyzes the dynamics of the front separating the responding/unresponding domains, in other words, the dynamics of the tension propagation along the backbone.

At time $t$, the positions of the front and the dragged segment are $x=-X(t)$ and $x=l(t)$, respectively (see Fig.~\ref{transient_image}). 
The responding part composing $M(t)$ segments in the range $x \in [-X(t), l(t)]$ assumes steady-state conformations moving with the representative velocity $\simeq V(t)$, for which the result obtained in Sec.~\ref{Scaling_steady} can be applied with the suitable redefinition of the variables, i.e., $L \rightarrow l+X$, $N \rightarrow M$. 
From eqs.~(\ref{f-L-V_2}) and (\ref{f-V-N}), we obtain
\begin{eqnarray}
\Bigl[ l(t)+X(t) \Bigr] V(t) &\simeq& \frac{a^2}{\tau_0} \left( \frac{f a}{k_BT}\right)^{\alpha}
\label{l-X-V-f} \\
M(t)V(t) &\simeq& \frac{a}{\tau_0}  \left(\frac{fa}{k_BT} \right)^{\alpha - \beta}. 
\label{M-V-f}
\end{eqnarray}
The moving distance of the driving site is 
\begin{eqnarray}
l(t) = \int_0^t {\rm d}s V(s)
\label{l-V}
\end{eqnarray}
Recalling that the chain is essentially unaffected behind the front, one can notice that the front position $X(t)$ and the segment label $M(t)$ (counting from the pulled end) at the front satisfies the relation
\begin{eqnarray}
X(t) \simeq a M^{\nu}.
\label{X-M}
\end{eqnarray}
Equations~(\ref{l-X-V-f})--(\ref{X-M}) constitute a closed set of the relations of $V(t), M(t), R(t)$.
The solution can be written in the form
\begin{eqnarray}
Q(V(t))-Q(V_0)= \frac{t}{\tau_0}
\label{Q_eq}
\end{eqnarray}
where we have defined the function
\begin{eqnarray}
Q(V)&=& \left(\frac{a}{\tau_0 V}\right)^2\left( \frac{fa}{k_BT}\right)^{\alpha} \left[ 1- \left(\frac{V_{\rm ss}}{V}\right)^{- \beta \nu}\left( \frac{f R}{k_BT}\right)^{-\beta}\right] \nonumber \\
&\simeq&  \left(\frac{a}{\tau_0 V}\right)^2\left( \frac{fa}{k_BT}\right)^{\alpha}
\label{Q-t_1}
\end{eqnarray}
with $V_{\rm ss}$ being the steady state velocity given by eq.~(\ref{f-V-N}).
The last near-equality in eq.~(\ref{Q-t_1}) holds for the situation of our interest $f \gg k_BT/R$.

Except for the very initial period, eq.~(\ref{Q-t_1}) can be approximated as  $Q(V(t)) \simeq t/\tau_0$ which can be arranged into
\begin{eqnarray}
V(t)\simeq V_{\rm ss}   \left(\frac{t}{\tau_{\rm tr}}\right)^{-1/2} \simeq  \frac{a}{\tau_0} \left( \frac{fa}{k_BT}\right)^{\alpha/2 } \left(
\frac{t}{\tau_0}\right)^{-1/2}  \label{V_t}
\label{transient_time_V}
\end{eqnarray}
where the characteristic time of the transient period can be deduced from the condition $\tau_{\rm tr}/\tau_0  \simeq Q(V(\tau_{\rm tr})) =  Q(V_{\rm ss}) $ so that
\begin{eqnarray}
\tau_{\rm tr} \simeq \tau_0 N^2 \left( \frac{fa}{k_BT}\right)^{\omega} 
\label{V_transient}
\end{eqnarray}
with the exponent 
\begin{eqnarray}
\omega \equiv 2 \beta -\alpha =\left\{
\begin{array}{ll}
              (2/\nu) -z &   ( k_{{\rm B}}T/R < f < k_{{\rm B}}T/a) \\
               -1 &  (f > k_{{\rm B}}T/a) 
           \end{array}
        \right.  \label{omega}
\end{eqnarray}
Note that in the moderately driven regime ($k_BT/R < f < k_BT/a$), eq.~(\ref{V_transient}) can be rewritten as $\tau_{\rm tr} \simeq \tau_{\rm eq} (fR/k_BT)^{\omega}$.
At the threshold force $f \simeq k_BT/R$, we see the crossover between the non-equilibrium and equilibrium dynamics, $\tau_{\rm tr} \simeq \tau_{\rm eq}$,  where $\tau_{\rm eq} \equiv \tau_0 (R/a)^z = \tau_0 N^{\nu z}$ is the longest relaxation time of the equilibrium coil. 

Combining eq.~(\ref{transient_time_V}) with eq.~(\ref{M-V-f}), we obtain the time evolution of the propagation front as
\begin{eqnarray}
M &\simeq& \left( \frac{f a}{k_BT}\right)^{-\omega/2 } \left( \frac{t}{\tau_0} \right)^{1/2}.
\label{propagation_front}
\end{eqnarray}
We thus find that the tension always propagates diffusively along the chain backbone with the {\it force dependent amplitude}. The exponent $\omega$ determines the force dependence in the transient process (eqs.~(\ref{V_transient}) and~(\ref{propagation_front})). For Rouse chain in the trumpet regime ($\nu = 1/2$, $z=4$), we have $\omega = 0$, in agreement with the exact result obtained in Sec.~\ref{transient_Rouse}. With the excluded-volume effect included ($\nu > 1/2$) in the free-draining case, we have the force amplitude exponent $\omega = (1/\nu) -2  < 0$. Surprisingly, this exponent reverses the sign in the non-draining case; $\omega = (2/\nu)  - 3>0$ irrespective of the excluded-volume effect.
One can obtain a clue on this somewhat counterintuitive result by rewriting eq.~(\ref{V_transient}) as
\begin{eqnarray}
\tau_{\rm tr} \simeq \tau_0  \left[ \frac{N (fa/k_BT)^{\beta}}{N^{-1}(fa/k_BT)^{\alpha-\beta}}\right] \simeq  \left( \frac{L_{ss}}{V_{ss}}\right)
\label{V_transient_}
\end{eqnarray}
where we have utilized the dynamical equation of states, i.e., eqs.~(\ref{f-V-N}) and~(\ref{f-L-N}).
While both the steady-state extension $L_{ss}$ and the velocity $V_{ss}$ increases with $f$, their relative rate generally depends on the dissipation mechanism and the excluded-volume effect.

Under the stronger force ($f > k_{{\rm B}}T/a$), the chain is highly stretched so that the excluded-volume effect becomes irrelevant, and we would expect the exponent $\omega=-1$.
It may be worth noting that the initial velocity $V_0$ is given from the condition
$\Gamma[\xi_0 ]V_0 = f$;
with the initial blob size $\xi_0 \simeq k_BT/f$
\begin{eqnarray}
V_0 \simeq \frac{a}{\tau_0}\left( \frac{fa}{k_BT}\right)^{z-1}
\label{V_0}
\end{eqnarray}
This has the same scaling structure for the downstream blob size $\xi_{{\rm free}}$ in the steady state (eq.~(\ref{xi_free})). Just as $\xi_{{\rm free}}$ saturates to the segment size at large force $f > Nk_BT/a$, the initial blob size reaches the lower bound $ \xi_0 \simeq a$ at the threshold force $f \simeq k_BT/a$, thus, the initial velocity is modified as $V_0 \simeq (a/\tau_0) (fa/k_BT)$ for larger forces.

\section{Nonlinear diffusion equation}
\label{Diffusion_eq}
The above scaling predictions based on the two-phase formalism can be corroborated by a semi-quantitative argument that considers a viscous dynamics of the local segment density in a single polymer chain. 
We will here outline this approach and derive a nonlinear diffusion equation that can describe the spatial profile of chain deformations in the transient process. 
Analytic solutions are available to our nonlinear diffusion equation, from which a number of the scaling predictions are confirmed. 

Take one dimensional coordinate $x$ along a pulling direction.
The cross section of the polymer at $x$ is $\xi(x,t)$, the velocity field of the segments averaged over this cross section is mainly along $x$ direction, which we denote as $v(x,t)$. 
The mass conservation averaged over the cross section at $x$ can be written as
\begin{eqnarray}
 \frac{\partial}{\partial t}(\rho\xi^2)+\frac{\partial}{\partial x}(\rho \xi^2 v) &=& 0,
 \label{sup:mass-cons}
\end{eqnarray}
The segment density $\rho$ satisfies $\rho \xi^3 \simeq a^3 g$, where $g$ is the number of segments within a blob of size $\xi$, and could be given due to the Flory relation as
$\xi \simeq a g^{\nu}$, from which we obtain $\rho \xi^2 \simeq a^2 (\xi/a)^{\beta}$. 
The local force balance along $x$-direction may be written as
\begin{eqnarray}
 \frac{\partial T}{\partial x}-\frac{\Gamma}{\xi} v &=& 0,
 \label{sup:force-balance}
\end{eqnarray}
where $T(x,t)$ is an internal tension in the polymer chain (averaged over the cross section at $x$), and $\Gamma$ is given in eq.~(\ref{eq:Gamma}).
The Pincus rule suggests $T \simeq k_BT/\xi$.
The boundary condition at the pulling end, $x=l(t)$, is $T(l,t)=f$, or using $T \simeq k_BT/\xi$, it gives on the scaling level the boundary condition for $\xi$ as
\begin{eqnarray}
 \xi(l,t) &\simeq& \frac{k_BT}{f}.
 \label{sup:BC2}
\end{eqnarray}
From Eqs.~(\ref{sup:mass-cons}) and (\ref{sup:force-balance}), we obtain
\begin{eqnarray}
 v(x,t) &\simeq& \frac{k_BT}{\gamma}\left(\frac{a}{\xi}\right)^{z-3}\frac{\partial}{\partial x}\left(\frac{a}{\xi}\right),
 \label{sup:v}
\end{eqnarray}
Substituting this into eq.~(\ref{sup:mass-cons}) and introducing the line segment density $\phi = ag/\xi \simeq (\xi/a)^{\beta}$, we arrive at the nonlinear diffusion equation
\begin{eqnarray}
 \frac{\partial \phi}{\partial t} &=& \kappa\frac{\partial}{\partial x}\left[\phi^{-\alpha/\beta}\frac{\partial \phi}{\partial x}\right],
 \label{sup:diffusion-2}
\end{eqnarray}
where the diffusion constant $\kappa$ includes unknown numerical factors which ensure the equality in Eq.~(\ref{sup:diffusion-2}), i.e., $\kappa=(\mbox{const}) \times \beta^{-1}k_BT/\gamma$.

{\it Solution and scaling behavior--}
Equation~(\ref{sup:diffusion-2}) or its analogue has recently been proposed to describe the polymer detachment kinetics from absorbing surface~\cite{Vilgis_2012} or the decompression/unfolding dynamics of an initially compacted polymer~\cite{PRL_Sakaue_2009}.
More traditionally, this type of nonlinear diffusion equation is often encountered in the ecology, i.e., insect and animal dispersal and invasion~\cite{Murray}. It also describes other various phenomena, sometimes called as the fast ($-\alpha/\beta<0$) or the slow ($-\alpha/\beta>0$) diffusion process, the relevance of which ranges from plasma physics, kinetic theory of gases to the fluid transport in porous medium.

The scaling relation for the dynamics of pulled end simply follows from the self-similar property of the system. Let $l(\tau)$ and $\phi_0(l(\tau),\tau)$ be the position of the pulled end at time $\tau$ and the segment line density there.
The self-similarity implies that eq.~(\ref{sup:diffusion-2}) is invariant under the scale transformation $t \rightarrow \tau {\tilde t}$, $x \rightarrow l(t) {\tilde x}$ and $\phi \rightarrow \phi_0 {\tilde \phi}$.
This leads to the relation $l^2 \simeq \tau \kappa \phi_0^{-\alpha/\beta}$
This, together with the boundary condition (eq~(\ref{sup:BC2})) $\phi_0 = (fa/k_BT)^{-\beta}$, leads to the time evolution of the pulled end $l(\tau)/a \simeq (\tau/a)^{1/2}(fa/k_BT )^{\alpha/2}$, which is precisely in accord with eq.~(\ref{V_t}).

To obtain a solution, it is convenient to introducing the variable $h= (\xi/a)^{-\alpha}$. Then eq.~(\ref{sup:diffusion-2}) can be rewritten as
\begin{eqnarray}
 \frac{\partial h}{\partial t} &=& -\frac{\beta\kappa}{\alpha}\left(\frac{\partial h}{\partial x}\right)^2+\kappa h \frac{\partial^2 h}{\partial x^2},
 \label{sup:diffusion-h}
\end{eqnarray}
Analytic forms of self-similar solutions for eq.~(\ref{sup:diffusion-h}) are known for the case $\alpha/\beta < 2$~\cite{Barenblatt-PNAS-2000}.
The special solution that can satisfy the boundary condition (\ref{sup:BC2}) is
$ h(x,t) = A(t)+B(t)(x/a)^2$, where 
\begin{eqnarray}
A(t) &=& A_0 \ (1+ \omega_0 t)^{\alpha/(2\beta-\alpha)} \label{sol:A}\\
B(t) &=& B_0\ (1+\omega_0 t)^{-1},  \label{sol:B}
\end{eqnarray}
with $\omega_0= 2B_0 \kappa a^{-2} [(2 \beta/\alpha) -1] \simeq \tau_0^{-1}$.

The polymer shape in the transient is thus found to be 
\begin{eqnarray}
 \xi(x,t) &=&  \xi(0,t) \left[1+\frac{x^2}{\lambda^2(t)}\right]^{-1/\alpha},
 \label{sup:xi-sol}
\end{eqnarray}
where $\xi(0,t)/a =  [A(t)]^{-1/\alpha}$,   and
$\lambda(t)/a = [A(t)/B(t)]^{1/2} \simeq [1+ \omega_0 t]^{\beta/(2\beta-\alpha)}$.
From this and eq.~(\ref{sup:v}), the velocity field is obtained as
\begin{eqnarray}
v(x,t) \simeq B(t) \frac{x}{\tau_0}.
\label{v_profile}
\end{eqnarray}
This allows us to write the profile as
\begin{eqnarray}
\xi(x,t)/a \simeq  [B(t) (x/a)^2]^{-1/\alpha} \simeq \left( \frac{\tau_0 v(x,t)}{a^2} x\right)^{-1/\alpha}
\label{transient_profile}
\end{eqnarray}
which is valid in the region $x \gg \lambda$.
This coincides with the form of steady state profile given in eq.~(\ref{steady_profile}).

The boundary condition, eq.~(\ref{sup:BC2}), requires $A(t)+B(t)(l(t)/a)^2 \simeq (fa/k_BT)^{\alpha}$, from which we obtain the time evolution of the position of the pulling end $l(t)$.
In particular, when $(fa/k_BT)^{\alpha} \gg A(t)$, that is, when 
\begin{eqnarray}
 t/\tau_0 &\ll& \left(\frac{fR}{k_BT}\right)^{(2\beta -\alpha)},
 \label{sup:cond-t}
\end{eqnarray}
where $R \simeq \xi(0,0)$ is a typical equilibrium size of the polymer chain, we can find the scaling law for $l(t)$ given by
\begin{eqnarray}
 l(t) &\simeq& a\left(\frac{fa}{k_BT}\right)^{\alpha/2}\left(\frac{t}{\tau_0}\right)^{1/2},
 \label{sup:l(t)}
\end{eqnarray}
which again accords with eq.~(\ref{V_t}).
This scaling behavior is expected to be observed during the time region given by $1  \ll t/\tau_0 \ll (fR/k_BT)^{(2\beta -\alpha)} \simeq \tau_{\rm tr}/\tau_{\rm eq}$.
There will be a substantial time range satisfying this, given the nonequilibrium regime $fR/k_BT \gg 1$ under consideration.

As is evident from the time dependency of $\lambda$, the method of present analysis yields a physically meaningful solution when $\alpha/\beta < 2$. This condition is always satisfied for the non-draining dynamics ($z=3$), but not for the free-draining dynamics ($z=(2\nu + 1)/\nu$). In this regard, it is interesting to note a recent study on the polymer detachment kinetics, which demonstrates $l(t) \sim t^{1/2}$ scaling for the pulled end even when $\alpha/\beta > 2$~\cite{Vilgis_2012}. Although the adopted linearizing approximation of eq.~(\ref{sup:diffusion-2}) in ref.~\cite{Vilgis_2012} limits the range of their analysis to the weak force case only, it provides a further support for the scaling analysis in Sec.~\ref{transient_Scaling}. 
We also note that the point $2\beta=\alpha$ corresponds to the Rouse chain, i.e., $z=4$, $\nu=1/2$ and coincides with the condition $\omega=0$ at which the force exponent reverses its sign (eq.~(\ref{omega})). The Rouse chain is exactly analyzed using the Gaussian model in Secs.~\ref{Rouse_steady} and~\ref{transient_Rouse} and the results are already confirmed to agree with scaling predictions in Sec.~\ref{transient_Scaling}.

Lastly, we point out that the results obtained in this section shed some light on the two-phase formalism developed in Sec.~\ref{transient_Scaling}. The linear velocity profile eq.~(\ref{v_profile}) leads to the average velocity of the moving domain $v_{ave} = \int_{x_{tail}}^{x_{head}} v(x) dx /(x_{head}-x_{tail}) = (v_{head}-v_{tail})/2$, where subscripts ``head" and ``tail" indicate the beginning and the end points of the moving domain. Given $v_{tail} \ll v_{head}$, it is natural to set the representative velocity as $V \simeq v_{head}$, which corresponds to our equation~(\ref{l-V}) in Sec.~\ref{transient_Scaling}. The spatial profile of the moving domain in the transient process is sharper than that in the steady-state, i.e., compare eqs.~(\ref{transient_profile}) and~(\ref{steady_profile}) because of the velocity gradient in the moving domain in the former situation (eq.~(\ref{v_profile})).

\section{Summary and Perspectives}
\label{summary}
We have presented a comprehensive analysis for the dynamics of a dragged polymer, a paradigm for the nonequilibrium response of polymer chains.
Both the steady-state and transient properties have been clarified by means of the exact solution of Rouse model and the scaling approach.
The latter allows us to establish the nonlinear extension/friction laws in steady-state and the tension propagation law during the transient process in general form, which involves the Rouse model result in a particular case.

The present set-up of dragging chain could be realized by manipulation techniques using optical/magnetic tweezers or atomic force microscopy (AFM). 
Here, there are two basic protocols for the manipulation. (i) A constant force is applied to the end segment, and (ii) the end segment is dragged in a constant velocity. This latter case is realized by fixing the end segment in space and then moving the stage in a constant speed.
During the transient period the friction constant increases with time, and our analysis indicates that the time evolutions of the dynamic friction are generally different in these two ensembles.
In the constant force ensemble, the friction constant grows as
\begin{eqnarray}
\Gamma_f(t) \equiv  \frac{f}{V(t)} \simeq \gamma \left( \frac{fa}{k_BT}\right)^{\delta_f}\left( \frac{t}{\tau_0}\right)^{\delta_{f,t}}
\end{eqnarray}
with exponents $\delta_f = 1-\alpha/2$ and $\delta_{f,t} = 1/2$. On the other hand, in the constant velocity ensemble, then, the evolution of the friction constant is
\begin{eqnarray}
\Gamma_V(t) \equiv  \frac{f(t)}{V} \simeq \gamma \left( \frac{\tau_0 V}{a}\right)^{\delta_V}\left( \frac{t}{\tau_0}\right)^{\delta_{V,t}}
\end{eqnarray}
with $\delta_V =(2/\alpha)-1$ and $\delta_{V,t}=1/\alpha$.

Transcribing back to the original dynamical exponent in the trumpet regime $k_{{\rm B}}T/R < f < k_{{\rm B}}T/a$, we have $\delta_f =2-(z/2)$, $\delta_V=(4-z)/(z-2)$ and $\delta_{V,t}=1/(z-2)$.   For a chain with ideal statistics $\nu=1/2$ in the free-draining limit $z=(1+2\nu)/\nu=4$, there is no distinction between these two expressions, $\Gamma_f(t) = \Gamma_V(t) \simeq \gamma (t/\tau_0)^{1/2}$, which coincides with the result obtained for a Rouse model in eq.~(\ref{Rouse_t_f_M_V}).
In other cases, hydrodynamic and/or excluded-volume interactions give rise to distinct dynamical behaviors for each of two ensembles characterized by nonzero $\delta_f$, $\delta_V$ as well as the difference in $\delta_{f,t}$ and $\delta_{V,t}$. In particular, (i) in the non-draining approximation ($z=3$), $\delta_f = 1/2$, $\delta_V = 1$ and $\delta_{f,t}=1/2$, $\delta_{V,t}=1$; (ii) in the free-draining approximation ($z=(1+2\nu)/\nu$), $\delta_f=1-(1/2\nu)$, $\delta_V=2\nu -1$ and $\delta_{f,t}=1/2$, $\delta_{V,t}=\nu$. In both cases, the time evolution are different depending on the ensemble. 
For stronger force $k_{{\rm B}}T/a < f$, the chain is highly stretched, thus the result becomes insensitive to excluded-volume/hydrodynamic effects. Substituting $\alpha=1$ and $\beta=0$, we find $\delta_f = \delta_{f,t}= 1/2$ and $\delta_V = \delta_{V,t}=1$. In usual experiments in aqueous media, where the non-draining approximation is expected to be valid (neglecting a logarithmic correction), we thus predict $\delta_f = \delta_{f,t}= 1/2$ and $\delta_V = \delta_{V,t}=1$ in the entire force range $f>k_{{\rm B}}T/R$.
We have to keep in mind, however, that a very long chain would be required to obtain clear signals by avoiding the effect due to a micron-sized trapping bead in tweezers experiment and the cantilever in AFM. Giant DNAs and chromosomes (appropriately treated if necessary) might be good candidates to test our predictions.

Closely related to the present analysis is the dynamics of polymer translocation through a narrow pore, on which there have been numerous recent researches~\cite{JPhys_Milchev_2011}. It has been proposed and verified that the dynamics of driven translocation is described as the sequential nonequilibrium process associated with the tension propagation along the chain~\cite{PRE_Sakaue_2008,PRE_Sakaue_2010} akin to the transient process of the dragged chain. Only distinction lies in the way how the driving force is exerted, i.e., unlike the end-pulling for the dragged chain (eq.~(\ref{f_Rouse_steady})), the force is acting only on the segment inside the pore which is fixed in space. Since the polymer exhibits a biased Brownian motion, the segment under the action of the driving force changes with time. This makes a trouble in the implementation of the force term in Rouse model, and also causes some subtlety in the description of the moving domain in the scaling approach~\cite{JPhyChem_Grosberg_2011, EPJE_Saito_2011}. Nonetheless, the basic physics can be well captured by the two-phase formalism, which can be developed along the same line with the present problem. We expect that a list of related examples would build up by adding phenomena in biology and polymer science, where the notion of the nonequilibrium response plays an important role.

\acknowledgements
This work was supported by the JSPS Core-to-Core Program
``International research network for non-equilibrium dynamics of soft matter".
H. W. acknowledges the financial support from MEXT of Japan (Grant in Aid, No.22740274).


\begin{thebibliography}{21}


\bibitem{deGennesBook}
	{ P.-G.~de~Gennes,
  {\it Scaling Concepts in Polymer Physics}
  (Cornell University Press, Ithaca, 1979).}
  
\bibitem{GrosbergBook}
	{ A.~Y. Grosberg and A.~R. Khokhlov,
  {\it Statistical Physics of Macromolecules}
  (AIP Press, New York, 1994).}
  
\bibitem{Macromolecules_Pincus_1976}
	{ P. Pincus,
  Macromolecules \textbf{9}, 386 (1976).}
  
\bibitem{Macromol_Marko_1995}
	{ J.~F.~Marko and E.~D. Siggia, 
	Macromolecules {\bf 28}, 8759 (1995).}
	
	
\bibitem{EPL_Brochard_1993}
	{F.~Brochard-Wyart,
  Europhys. Lett. {\bf 23}, 105 (1993).}
  

\bibitem{EPL_Brochard_1995}
	{F.~Brochard-Wyart,
  Europhys. Lett. {\bf 30}, 387 (1995).}
	
\bibitem{Science_Chu_1995}
	{ T.~T. Perkins, D.~E. Smith, R.~G. Larson and S. Chu, 
	Science {\bf 268}, 83 (1995).}

\bibitem{PRE_Larson_Chu_1997}
	{ R.~G. Larson, T.~T. Perkins, D.~E. Smith, and S. Chu, 
	Phys. Rev. E {\bf 55}, 1794 (1997).}

\bibitem{Macro_Netz_2008}
  {X.~Schlagberger and R.R.~Netz,
  Macromolecules \textbf{41}, 1861 (2008).}  
	
\bibitem{EPL_Brochard_1994}
	{F.~Brochard-Wyart, H. Hervet and P. Pincus,
  Europhys. Lett. {\bf 26}, 511 (1994).}
	
  
\bibitem{Seifert_1996}
   {U. Seifert, W. Wintz and P. Nelson,
  Phys. Rev. Lett. {\bf 77}, 5389 (1996).}
  

\bibitem{PRE_Sakaue_2008}
	{ T.~Sakaue,
  Phys. Rev. E \textbf{76}, 021803 (2007).}  
  
\bibitem{PRE_Sakaue_2010}
	{ T.~Sakaue,
  Phys. Rev. E {\bf 81}, 041808 (2010).} 
  
\bibitem{EPL_Sebastian_2011}
    { K.L.~Sebastian, V.G. Rostiashvili and T.A. Vilgis,
    Europhys. Lett. \textbf{95}, 48006 (2011).} 
  
\bibitem{Vilgis_2012}
  { J.~Paturej, A.~Milchev, V.G. Rostiashvili and T.A. Vilgis,
  arXiv:1203.1565v1}  

\bibitem{DoiEdwards}
	{ M. Doi and S.~F.~Edwards,
  {\it The Theory of Polymer Dynamics}
  (Clarendon Press, Oxford, 1986).}
  
\bibitem{Barenblatt-PNAS-2000}
 {G. I. Barenblatt, M. Bertsch, A. E. Chertock and V. M. Protokishin, PNAS \textbf{97}, 9844 (2000).}
 
  


	





  
  


\bibitem{Murray}
	{J.D.~Murray,
   {\it Mathematical Biology}
  (Springer-Verlag, New York, 2002).}
  
\bibitem{PRL_Sakaue_2009}
	{T. Sakaue and N. Yoshinaga,
    Phys. Rev. Lett. \textbf{102}, 148302 (2009).}   

\bibitem{JPhys_Milchev_2011}
	{A. Milchev,
    J. Phys.: Condens. Matter \textbf{23}, 103101 (2011)  and references therein.}
    
\bibitem{JPhyChem_Grosberg_2011}
	{P. Rowghanian and A.Y. Grosberg,
    J. Phys. Chem. B \textbf{115}, 14127 (2011).}  

\bibitem{EPJE_Saito_2011}
	{T. Saito and T. Sakaue,
    Eur. Phys. J. E \textbf{34}, 135 (2011).}      






\end{thebibliography}
\end{document}